# FEASIBILITY OF OTR IMAGING OF NON-RELATIVISTIC IONS AT GSI*

A.H. Lumpkin, Fermi National Accelerator Laboratory, Batavia, IL 60510 U.S.A.

*Abstract*

The feasibility of using the optical transition radiation (OTR) generated as a 11.4- to 300-MeV/u ion beam passes through a single metal conducting plane for a minimally intercepting beam profile monitor for GSI/Darmstadt has been evaluated for the first time. Although these are non-relativistic beams, their beta and gamma values are similar to the 80-keV electron-beam imaging studies previously done on the CTF3 injector. With anticipated beam intensities of $10^9$ to $10^{11}$ particles per pulse and the predicted charge-squared dependence of OTR, the ion charge state becomes a critical factor for photon production. The OTR signal from the ion charge integrated over the video field time should be comparable to or larger than the CTF3 electron case. These signal strengths will allow a series of experiments to be done that should further elucidate the working regime of this technique.

## INTRODUCTION

The characterizations of the ion beams used for the various physics experiments at the Gesellschaft fur Schwerionenforschung (GSI) in Darmstadt, Germany is an ongoing interest. The basic beam profiling has been done with scintillator screens, secondary emission monitors (SEMs), a beam scraper and downstream detector, or beam-induced gas fluorescence (BIF) [1-3]. In the case of scintillators, the beam-image size variations that have been observed during bombardment are attributed to the temperature increases and actual ionizing radiation damage to the intrinsic properties of the scintillator materials used in a beam-intercepting mode [1,3]. As an alternative to the scintillators, I have evaluated the feasibility of using optical transition radiation (OTR) monitoring [4] as a minimally intercepting means to image the beam size of the 11.4- to 300-MeV/u beams with up to $10^{11}$ particles per pulse as they pass through a single conducting plane. Generally, the scintillator screens have 500 times more light emission than OTR from relativistic beams into a light collecting lens system, and for low beta the very wide angular distribution lobes present an even bigger photon-collection issue. However, previous experiments were successfully done using near-field imaging on the low-beta (0.63) 80-keV electron beam in the CLIC Test Facility 3 (CTF3) injector test stand with about $10^{11}$ electrons in 4 ns using an ICCD [5]. In addition, we note that intense proton beams (albeit relativistic) with $10^{13}$ ppp delivered in 10 µs and in a 1-mm radius beam spot have been imaged with thin converter screens such as aluminized Kapton, Ti foils, or Al foils. The first type survived in the proton beams for months in the beamline just before the target for neutrino production [6]. With the high energy depositions of ions, such thin foils may be a way to circumvent the heating issues as seen in the thicker and less heat conducting scintillator materials. Since OTR is a surface phenomenon, one could potentially use for the GSI ion beams the 1000-Angstrom Al coatings on 6-µm thick Kapton substrates such as used at FNAL. Finally, it is proposed that the OTR photon number will be dependent on the square of the ion charge state, and this factor will compensate sufficiently for the low beta and the wide angular distributions to allow imaging the ion beams with standard, intensified CCD cameras.

## BACKGROUND ASPECTS

*GSI Complex*

The GSI complex is designed to accelerate ions from hydrogen to Uranium with energies of 3.8 to 11.4 MeV/u in a universal rf linear accelerator (UNILAC). Additional accelerations up to 300 MeV/u are possible in the rapid cycling synchrotron [3]. A schematic of the facility is shown in Fig. 1. A diagnostics development beam line station is presently located on beamline X2, and it has been used for a number of beam profiling tests in the past. It is a potential location for the first tests of OTR with ions since it already has the test chamber, basic screen holder, port at 90 degrees to the beam direction, and imaging options with CCDs or ICCDs [7].

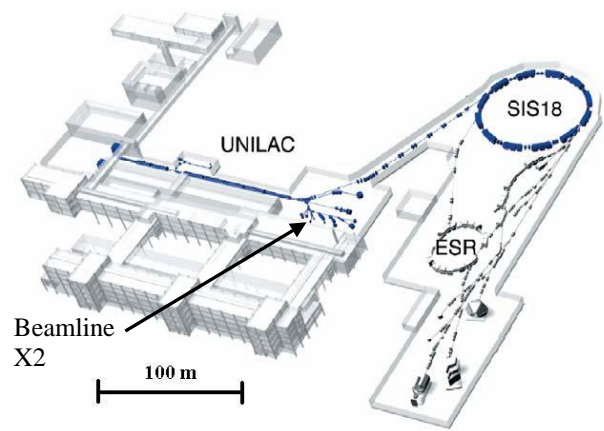

Figure 1: Schematic of the GSI indicating the UNILAC, test beamline X2, and Synchrotron (SIS18) as shown in ref. [3].

_________________
*Work supported by the U.S. Department of Energy, Office of Science, Office of High Energy Physics, under Contract No. DE-AC02-06CH11357.

*OTR Background*

There is extensive experience with OTR imaging of relativistic electron beams [e.g. 8] since Wartski's initial tests and some with proton beams [6]. In these cases gamma was >10 and β=0.99. In the GSI ion and heavy-ion cases, we are dealing with non-relativistic beams with β = 0.15-0.65. Since photon production does go as beta squared for low beta beams, one does need a modest beta value at least. We propose that the reference case of 80-keV electron beam imaging done at CTF3 provides a good starting point for estimating the feasibility of imaging the ion beams. In this case the angular distribution patterns are quite different from the relativistic cases, and are as shown in the Figs. 1,4,6 of ref. [5] and measured in that paper.

By integrating over 60 nC in a 4-ns pulse, they did obtain usable images of the beam spot and even tracked the linearity over a range of charges from 20 to 60 nC (the latter is 3 x $10^{10}$ particles). With the low calculated photon yield of 2 x $10^{-6}$ visible photons per electron, an ICCD was needed to image the estimated 7 x $10^5$ photons signal level in the 20-mrad collection angle [5]. A 1-mm thick Al screen at 20 degrees to the beam direction generated the backward OTR that had its stronger lobe emitted over 40 degrees centered at ~80 degrees to the beam direction as illustrated in Fig. 2 (from ref. [5]).

Recently, at Fermilab an electron gun generating 60-keV electrons at 1 mA has been commissioned. Initial tests with a polished stainless steel OTR screen have shown that a pulse train structure can be used so that temperature effects and blackbody radiation are mitigated [9]. Then with about 1.2 µC of integrated charge, the 110-µm (σ) beam spot was imaged with only a 10-bit Firewire CCD camera. The two screens were at 15 and 20 degrees to the beam direction with the camera at 90 degrees as guided by the CTF3 results and CERN's Mathcad-based OTR code. Linear polarizers and bamdpass filters were used to distinguish the sources, as well as using the promptness of the OTR mechanism.

Ginzburg and Tsytovich [10] considered a non-relativistic charge q moving from vacuum to an ideal conductor with v << c. They write the spectral energy density integrated over all angles as:

$$W(\omega) = \frac{4}{3\pi} \frac{q^2}{c} \left(\frac{v}{c}\right)^2 N \quad (1)$$

where ω = radiation frequency, v = particle velocity, c=speed of light, q = particle charge, N is the particle number.

One hypothesizes q$^2$ = (Ze)$^2$, where Z is the ion charge state and e is the magnitude of electron charge. Since we measure light intensity *I*, this should be proportional to $|E_x|^2 + |E_y|^2$. The incoherent photon intensity is proportional to N, the number of particles.

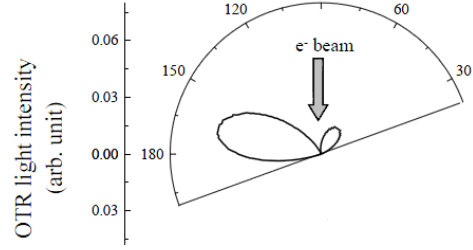

Figure 2: Calculation of the far-field backward OTR radiation from a single metal plane tilted at 20 degrees to the beam direction. (from ref. [5]).

## OTR FEASIBILITY CONSIDERATIONS

In Fig. 3 a schematic of the proposed experiment is shown with the key features of ion intensity, robust thin screen, screen angle for low beta, and the imaging system with an ICCD camera. The ICCD gating feature could also be used to select preferentially the prompt OTR versus any background sources in the scene that have longer emission time constants such as blackbody radiation from the heated up foil due to the ion bombardment or residual gas BIF sources.

The extension of the OTR imaging to the GSI ions is dependent on the β value, OTR photon yield, beam intensity, and ion charge state values. Table I shows a comparison of a few example ion cases with previous successful electron (non-relativistic [5] and relativistic [10]) and proton cases [6]. Potentially the higher charge states of heavy ions such as U$^{+28}$ and U$^{+73}$ compared to Ar$^{+10}$ would be advantageous. It is assumed an aluminized Kapton screen or carbon foil will survive the ion beams, and an ICCD can be used for imaging. Simplistically, it appears that the Ar ion case may be on the low signal side of feasibility, but that any ion beam with charge states greater than 10 and intensities over $10^{10}$ ppp in 1-2 mm spot size could be potentially imaged.

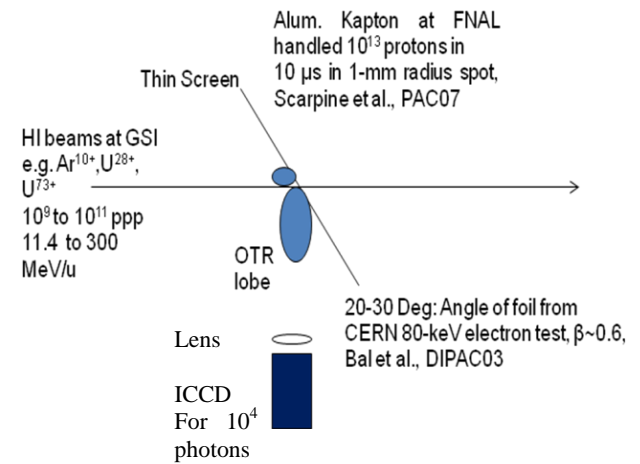

Figure 3: Schematic of the OTR experiment for ions with the thin screen, the low-beta angle of 20-30 degrees to the beam direction of the screen to obtain signal near 90 degrees to the beam direction, and the ICCD.

TABLE I. Comparison of various particle beam cases with photon numbers estimated. The charge state (Z), Yield (Y), Multiplier Factor (Mult.) for $\beta^2$ and $Z^2$ referenced to the 80-keV electron case, final estimated photon number, and whether a CCD or CID was used with intensifier (Int.) are given in addition to particle parameters.

| Particle | E (MeV) | Z | $\beta$ | $\gamma$ | Y(ph/e) | N | Mult. | Photon No. | CCD |
|---|---|---|---|---|---|---|---|---|---|
| $e^-$ | 0.080 | 1 | 0.63 | 1.15 | $2 \times 10^{-6}$ | $4 \times 10^{11}$ | 1 | $7 \times 10^5$ | Int. |
| $e^-$ | 150 | 1 | 0.99 | 300 | $2 \times 10^{-3}$ | $6 \times 10^9$ | - | $1.2 \times 10^7$ | yes |
| $p^+$ | $120 \times 10^3$ | 1 | 0.99 | 129 | $10^{-3}$ | $10^{11}$ | - | $10^8$ | CID |
|  | MeV/u |  |  |  |  |  |  |  |  |
| Ar | 11.4 | 10 | 0.15 | 1.01 | $10^{-6}$ | $10^{10}$ | 5.3 | $5 \times 10^4$ | *Int. |
| U | 11.4 | 28 | 0.15 | 1.01 | $10^{-6}$ | $10^{11}$ | 42 | $4 \times 10^6$ | *Int. |
| U | 300 | 73 | 0.65 | 1.21 | $10^{-6}$ | $10^9$ | 5329 | $5 \times 10^6$ | *Int. |

*The image intensifer is proposed for these ion cases for both the gain and the gating feature.

## SUMMARY

In summary, the feasibility of applying the backward OTR near-field, beam-size monitor technique to the ion beams in the GSI experimental halls looks very promising. It is more than a *gedanken* experiment. Actual tests in one of the beamlines should help to identify any unexpected background source levels or other operational issues. The next step would be to install several stations in the transport lines to provide the online feedback for beam size (emittance). One also anticipates that the scaling to the more intense and energetic beams planned in the Facility for Antiproton and Ion Research (FAIR) will be straightforward (subject to foil survivability) *if* one performs the parameter scaling validations successfully at 11.4 to 300 MeV/u. In addition, the potential for applications to other laboratories' intense ion beams should be considered.

## ACKNOWLEDGMENTS

The author acknowledges discussions with B. Walasek and P. Forck of GSI on ion beam intensities, beam energies, and diagnostic issues at GSI and with D. Rule (NSWC) and E. Bravin (CERN) on OTR aspects.